\documentclass{kapproc} 
\usepackage{graphicx}
\normallatexbib
\begin{document}

\articletitle{Realization of an $N$-shaped $IVC$ of nanoscale
metallic junctions using the antiferromagnetic transition. }

\author{Yu. G. Naidyuk, K. Gloos$^*$, I. K. Yanson}

\affil{B.Verkin Institute for Low Temperature Physics and
Engineering, National Academy  of Sciences of Ukraine, 47 Lenin
Ave., 61103, Kharkiv, Ukraine}

\affil{$^*$Nano-Science Center, Niels Bohr Institute fAFG,
Universitetsparken 5, DK-2100 Copenhagen, Denmark}

\email{naidyuk@ilt.kharkov.ua}

\chaptitlerunninghead{Realization of an $N$-shaped $IVC$ in
nanoscale metallic junctions}

\begin{abstract}
We have observed at low temperatures ($\leq 8$\,K) hysteretic
$I(V)$ characteristics for sub-$\mu$m ($\sim$200\,nm) metallic
break-junctions based on the heavy-fermion compound UPd$_2$Al$_3$.
Degrading the quality of the contacts by in situ increasing the
local residual resistivity or temperature rise reduces the
hysteresis. We demonstrate that those hysteretic $I(V)$ curves can
be reproduced theoretically by assuming the constriction to be in
the thermal regime. Our calculations show that such anomalous
$I(V)$ curves are due to the sharp increase of $\rho(T)$ of
UPd$_2$Al$_3$ near the N\'{e}el temperature $T_N \simeq 14$\,K.
From this point of view each metal with similar $\rho(T)$ should
produce similar hysteretic $I(V)$ curves. As example we show
calculations for the rare-earth manganite
La$_{0.75}$Sr$_{0.25}$MnO$_3$, a system with colossal
magnetoresistance.  In this way we demonstrate that nano-sized
point contacts can be non-linear devices with $N$-shaped $I(V)$
characteristics, i.~e. with negative differential resistance, that
could serve like Esaki tunnel diodes or Gunn diodes as amplifiers,
generators, and switching units. Their characteristic response
time is estimated to be less than 1\,ns for the investigated
contacts.



\end{abstract}


Point-contact (PC) spectroscopy is widely used to study the
interaction of conduction electrons with elementary excitations or
quasiparticles in conducting solids \cite{Naidyuk}. On the other
hand, PC investigations can shed light on peculiarities of the
electronic transport in nano-scale devices at ultrahigh current
densities. The latter matters for mesoscopic or nanoscale physics,
and especially for applied research where electronic devices have
already reached this sub-$\mu$m scale.

Electron transport in nanostructures can be distinguished by
basically three different current regimes, depending on the
relationship between the elastic $l_{\rm el}$ and the inelastic
$l_{\rm in}$ mean free path of electrons,  and the constriction or
point-contact diameter $d$ (for a review see \cite{Naidyuk},
Chapter 3). The constrictions are either ballistic ($l_{\rm el},
l_{\rm in}  \gg d$), diffusive ($l_{\rm el}\ll d \ll \sqrt{l_{\rm
in} l_{\rm el}}$), or thermal ($l_{\rm el}, l_{\rm in}\ll d$).

{\it A priori} it  is not trivial to determine the specific
current regime, which is important to further characterize the
nano-object: first, because the nano-sized object needs a
well-defined geometry; second, to evaluate the electronic mean
free path, especially the inelastic one, can be rather
speculative. To investigate the non-linear $I(V)$ curves using PC
spectroscopy seems to be the best method to solve this problem.

\begin{figure}[b]
\centering\includegraphics[width=9cm,angle=0]{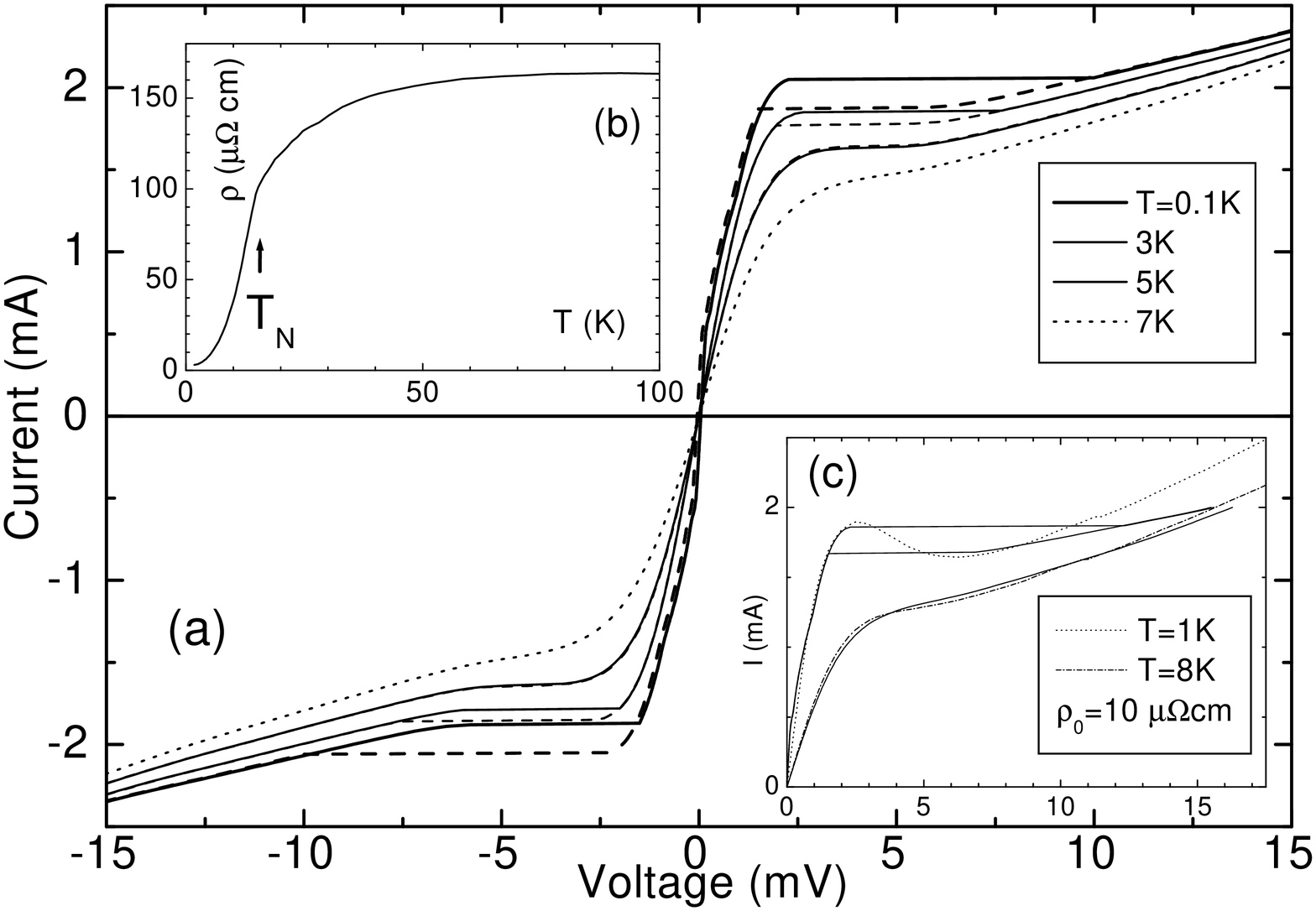}
\caption[]{$I(V)$ curves of a UPd$_2$Al$_3$ break junction with
$R_n=0.66\,\Omega$ at the indicated temperatures. Solid (dashed)
lines correspond to sweeps with increasing (decreasing) current.
The hysteretic loops become smaller when the temperature rises and
vanish above $\sim 5\,$K. Inset: (b) $\rho(T)$ of the bulk
compound \cite{Geibel}, (c) measured (solid) and calculated
(dashed) $I(V)$ curves at two temperatures. After
\cite{Naidyuk1}.} \label{updal}
\end{figure}

Here we present experiments on PCs between two pieces of the
heavy-fermion compound UPd$_2$Al$_3$, using the technique of
mechanically-cont\-rollable break junctions (see \cite{Naidyuk},
Chapter 4.1.5). UPd$_2$Al$_3$ becomes antiferromagnetic (AFM) at
$T_N \simeq$14\,K \cite{Geibel}. We have observed huge
non-linearities of the PC resistances and even hysteretic $I(V)$
characteristics (Fig.\ref{updal}). We derived the contact size and
the residual resistivity in the PC region as described in
\cite{Naidyuk1}. We found that the very short elastic mean-free
path in the constriction $l_{\rm el}\ll d$ points to at least the
diffusive regime of electron transport through the PC. Considering
also the small inelastic mean-free path in UPd$_2$Al$_3$,
reflected by the steep $\rho(T)$ rise with temperature around the
AFM transition in Fig.~\ref{updal}(b), we applied the thermal
model developed in Refs.~\cite{Verkin,Kulik}. In this case the
excess electron energy $eV$ dissipates within the constriction,
increasing the temperature inside the contact when a bias voltage
$V$ is applied.  As a result $I(V)$ is governed by the resistivity
$\rho (T)$ via \cite{Verkin,Kulik}
\begin{equation}
\label{IVT}
  I(V) = Vd \int_0^1  \frac{{\rm d}x}{\rho(T
\sqrt{1-x^2}~)},
\end{equation}
where the temperature $T$ in the center of the constriction is set
by $T^2 = T^2_{\rm bulk} + V^2/4L$ and $L$ is the Lorenz number.

The calculated $I(V)$ curve at $T=1$K in Fig.~\ref{updal}(c) has
maximum at around $2.5\,$mV, resulting in a hysteresis for up- and
downward sweeps when the junction is driven by a current source.
Figure \ref{updal}(c) shows that the theoretical $I(V)$ describe
well the experimental data, including the width of the hysteresis,
using $d = 200\,$nm and $\rho_0 = 10\,\mu\Omega$cm, where $\rho_0$
is the additional residual resistance: $\rho(T)=\rho_0 + \rho_{\rm
bulk}(T)$. These are the only two adjustable parameters which have
been derived independently from the measured contact resistance
$R(T)$ in \cite{Naidyuk1}. This agreement strongly supports our
interpretation that local thermal effects at the PC determine the
behavior of our UPd$_2$Al$_3$ break-junction conductivity.

The UPd$_2$Al$_3$ junctions presented here are non-linear devices.
Their $N$-shaped $I(V)$ characteristics have a negative
differential resistance at very high current densities up to
$5\times 10^{10}\,$A/m$^{2}$. Those devices could be applied -- in
principle -- like an Esaki tunnel diode or a Gunn diode
\cite{Esaki,Price} as amplifiers,  generators, or switching units.
Of practical interest is therefore the possible minimum response
time. We estimate \cite{Naidyuk1} a thermal relaxation time  $\tau
\approx 100\,$ps for a $d = 100\,$nm wide contact. This is three
orders of magnitude larger than for a standard tunnel diode, but
it could be reduced by using smaller contacts as long as they
remain in the thermal regime.

Obviously UPd$_2$Al$_3$ is not such a unique material for creating
$N$-shaped IVCs -- each metal with  a similar $\rho (T)$ should
also produce similar $I(V)$ characteristics. This can be expected
for many materials which order magnetically, since their
resistivity typically increases steeply when the magnetic order is
destroyed by thermal fluctuations.

\begin{figure}[t]
\centering\includegraphics[width=\textwidth,angle=0]{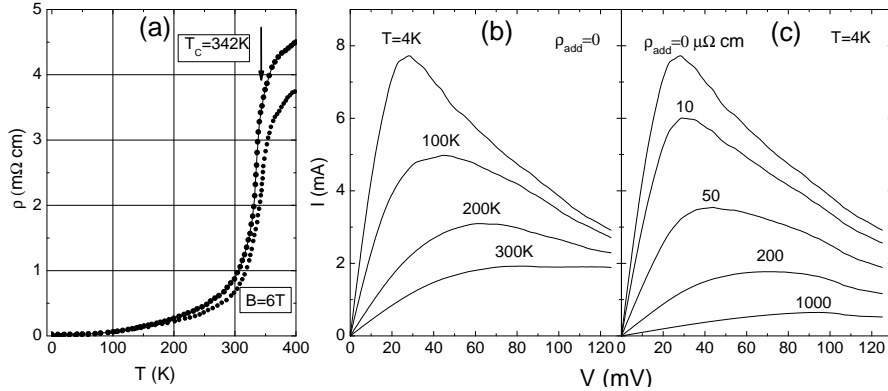}
\vspace{-4.5cm} \caption[]{ (a) Resistivity $\rho(T)$ of
La$_{0.75}$Sr$_{0.25}$MnO$_3$ at zero (upper curve) and $B$=6\,T
magnetic field according to \cite{Mitra}. (b) $I(V)$ calculated
using Eq.(\ref{IVT}) for $d$ = 100\,nm at the indicated
temperatures, and (c)  for different residual resistances at
$T$=4\,K. } \label{lsmo}
\end{figure}

An example for this behaviour is the well known rare-earth
manganite La$_{0.75}$Sr$_{0.25}$MnO$_3$, a system with colossal
magnetoresistance as shown in Fig.\ref{lsmo}(a). Indeed, the
calculated $I(V)$ curves in Fig.~\ref{lsmo}(b) are $N$-shaped up
to room temperature. Even enhancing disorder by increasing the
residual resistivity up to $\rho_{\rm add}=1000\,\mu\Omega$cm as
in Fig.~\ref{lsmo}(c) does not suppress the hysteresis at low
temperatures.

On the other hand, the position of the current maxima shift to
higher voltages with increasing temperature (or residual
resistivity), contrary to what has to be expected for a simple AFM
transition. Thus it would be very interesting to study $I(V)$ for
constrictions of this compound experimentally.

The partial support of the complex program "Nanosystems,
nanomaterials and nanotechnologies" of the National Academy of
Sciences of Ukraine are acknowledged.

{}

\end{document}